\documentclass[twocolumn,showpacs,preprintnumbers,superscriptaddress,amsmath,amssymb,PRL]{revtex4}
\usepackage{graphicx}
\usepackage{dcolumn}
\usepackage{float}
\usepackage{mathptmx, courier, pifont}
\usepackage[scaled=0.92]{helvet}
\usepackage[T1]{fontenc}
\usepackage{textcomp}
\usepackage{color}
\usepackage{booktabs}
\usepackage[dvipsnames]{xcolor}
\usepackage[colorlinks=true,urlcolor=Blue,linkcolor=Blue]{hyperref}
\usepackage[all]{hypcap}

\begin{document}
\title{Mass measurements for $T_{z}=-2$ $fp$-shell nuclei $^{40}$Ti, $^{44}$Cr, $^{46}$Mn, $^{48}$Fe, $^{50}$Co and $^{52}$Ni}
\author{C.~Y.~Fu}
\affiliation{CAS Key Laboratory of High Precision Nuclear Spectroscopy, Institute of Modern Physics, Chinese Academy of Sciences, Lanzhou 730000, People's Republic of China}
\author{Y.~H.~Zhang}\thanks{Corresponding author. Email address: yhzhang@impcas.ac.cn}
\affiliation{CAS Key Laboratory of High Precision Nuclear Spectroscopy, Institute of Modern Physics, Chinese Academy of Sciences, Lanzhou 730000, People's Republic of China}
\affiliation{University of Chinese Academy of Sciences, Beijing 100049, People's Republic of China}
\author{M.~Wang}\thanks{Corresponding author. Email address: wangm@impcas.ac.cn}
\affiliation{CAS Key Laboratory of High Precision Nuclear Spectroscopy, Institute of Modern Physics, Chinese Academy of Sciences, Lanzhou 730000, People's Republic of China}
\affiliation{University of Chinese Academy of Sciences, Beijing 100049, People's Republic of China}
\author{X.~H.~Zhou}
\affiliation{CAS Key Laboratory of High Precision Nuclear Spectroscopy, Institute of Modern Physics, Chinese Academy of Sciences, Lanzhou 730000, People's Republic of China}
\affiliation{University of Chinese Academy of Sciences, Beijing 100049, People's Republic of China}
\author{Yu.~A. Litvinov}\thanks{Corresponding author. Email address: y.litvinov@gsi.de}
\affiliation{CAS Key Laboratory of High Precision Nuclear Spectroscopy, Institute of Modern Physics, Chinese Academy of Sciences, Lanzhou 730000, People's Republic of China}
\affiliation{GSI Helmholtzzentrum f\"{u}r Schwerionenforschung,
	Planckstra{\ss}e 1, 64291 Darmstadt, Germany}
\author{K.~Blaum}
\affiliation{Max-Planck-Institut f\"{u}r Kernphysik, Saupfercheckweg 1, 69117 Heidelberg, Germany}
\author{H.~S.~Xu}
\affiliation{CAS Key Laboratory of High Precision Nuclear Spectroscopy, Institute of Modern Physics, Chinese Academy of Sciences, Lanzhou 730000, People's Republic of China}
\author{X.~Xu}
\affiliation{CAS Key Laboratory of High Precision Nuclear Spectroscopy, Institute of Modern Physics, Chinese Academy of Sciences, Lanzhou 730000, People's Republic of China}
\affiliation{School of physics, Xi'an Jiaotong University, Xi'an 710049, China}
\author{P.~Shuai}
\affiliation{CAS Key Laboratory of High Precision Nuclear Spectroscopy, Institute of Modern Physics, Chinese Academy of Sciences, Lanzhou 730000, People's Republic of China}
\author{Y.~H.~Lam}
\affiliation{CAS Key Laboratory of High Precision Nuclear Spectroscopy, Institute of Modern Physics, Chinese Academy of Sciences, Lanzhou 730000, People's Republic of China}
\affiliation{University of Chinese Academy of Sciences, Beijing 100049, People's Republic of China}
\author{R.~J.~Chen}
\affiliation{CAS Key Laboratory of High Precision Nuclear Spectroscopy, Institute of Modern Physics, Chinese Academy of Sciences, Lanzhou 730000, People's Republic of China}
\author{X.~L.~Yan}
\affiliation{CAS Key Laboratory of High Precision Nuclear Spectroscopy, Institute of Modern Physics, Chinese Academy of Sciences, Lanzhou 730000, People's Republic of China}
\author{X.~C.~Chen}
\affiliation{CAS Key Laboratory of High Precision Nuclear Spectroscopy, Institute of Modern Physics, Chinese Academy of Sciences, Lanzhou 730000, People's Republic of China}
\author{J.~J.~He}
\affiliation{Key Laboratory of Beam Technology of Ministry of Education, College of Nuclear Science and Technology, Beijing Normal University, Beijing 100875, People's Republic of China}
\affiliation{Beijing Radiation Center, Beijing 100875, People's Republic of China}
\author{S.~Kubono}
\affiliation{CAS Key Laboratory of High Precision Nuclear Spectroscopy, Institute of Modern Physics, Chinese Academy of Sciences, Lanzhou 730000, People's Republic of China}
\author{M.~Z.~Sun}
\affiliation{CAS Key Laboratory of High Precision Nuclear Spectroscopy, Institute of Modern Physics, Chinese Academy of Sciences, Lanzhou 730000, People's Republic of China}
\author{X.~L.~Tu}
\affiliation{CAS Key Laboratory of High Precision Nuclear Spectroscopy, Institute of Modern Physics, Chinese Academy of Sciences, Lanzhou 730000, People's Republic of China}
\affiliation{Max-Planck-Institut f\"{u}r Kernphysik, Saupfercheckweg 1, 69117 Heidelberg, Germany}
\author{Y.~M.~Xing}
\affiliation{CAS Key Laboratory of High Precision Nuclear Spectroscopy, Institute of Modern Physics, Chinese Academy of Sciences, Lanzhou 730000, People's Republic of China}
\author{Q.~Zeng}
\affiliation{CAS Key Laboratory of High Precision Nuclear Spectroscopy, Institute of Modern Physics, Chinese Academy of Sciences, Lanzhou 730000, People's Republic of China}
\affiliation{School of nuclear science and engineering, East China University of Technology, Nanchang 330013, People's Republic of China}
\author{X.~Zhou}
\affiliation{CAS Key Laboratory of High Precision Nuclear Spectroscopy, Institute of Modern Physics, Chinese Academy of Sciences, Lanzhou 730000, People's Republic of China}
\affiliation{University of Chinese Academy of Sciences, Beijing 100049, People's Republic of China}
\author{W.~L.~Zhan}
\affiliation{CAS Key Laboratory of High Precision Nuclear Spectroscopy, Institute of Modern Physics, Chinese Academy of Sciences, Lanzhou 730000, People's Republic of China}
\author{S.~Litvinov}
\affiliation{GSI Helmholtzzentrum f\"{u}r Schwerionenforschung,
	Planckstra{\ss}e 1, 64291 Darmstadt, Germany}
\author{G.~Audi}
\affiliation{CSNSM-IN2P3-CNRS, Universit\'{e} de Paris Sud, F-91405
	Orsay, France}
\author{T.~Uesaka}
\affiliation{RIKEN Nishina Center, RIKEN, Saitama 351-0198, Japan}
\author{T. Yamaguchi}
\affiliation{Department of Physics, Saitama University, Saitama 338-8570, Japan}
\author{A.~Ozawa}
\affiliation{Insititute of Physics, University of Tsukuba, Ibaraki 305-8571, Japan}
\author{B.~H.~Sun}
\affiliation{School of Physics, Beihang University, Beijing 100191, People's Republic of China}
\author{Y.~Sun}
\affiliation{School of Physics and Astronomy, Shanghai Jiao Tong University,
	Shanghai 200240, People's Republic of China}
\author{F.~R.~Xu}
\affiliation{State Key Laboratory of Nuclear Physics and Technology, School of Physics, Peking University, Beijing 100871, People's Republic of China}
\date{\today}
\begin{abstract}
By using isochronous mass spectrometry (IMS) at the experimental cooler storage ring CSRe, masses of short-lived $^{44}$Cr, $^{46}$Mn, $^{48}$Fe, $^{50}$Co and $^{52}$Ni were measured for the first time and the precision of the mass of $^{40}$Ti was improved by a factor of about 2.
Relative precisions of $\delta m/m=(1-2)\times$10$^{-6}$ have been achieved.
Details of the measurements and data analysis are described.
The obtained masses are compared with the Atomic-Mass Evaluation 2016 (AME$^{\prime}$16) and with theoretical model predictions.
The new mass data enable us to extract the higher order coefficients, $d$ and $e$, of the quartic form of the isobaric multiplet mass equation (IMME) for the $fp$-shell isospin quintets.
Unexpectedly large $d$- and $e$-values for $A=44$ quintet are found.
By re-visiting the previous experimental data on $\beta$-delayed protons from $^{44}$Cr decay,
it is suggested that the observed anomaly could be due to the misidentification of the $T=2$, $J^\pi=0^{+}$ isobaric analog state (IAS) in $^{44}$V.  

%
\end{abstract}

\maketitle
\section{Introduction}
Atomic masses are widely applied to investigations in many areas of subatomic physics ranging from nuclear structure and astrophysics to fundamental interactions and symmetries depending on the mass precision achieved~\cite{Blaum06,Lunney03}.
The evolution of nuclear shell structure, nucleon correlations and changes of deformation
are often studied through observing systematic trends
of one- and two-nucleon separation energies, which are deduced directly from the atomic masses involved~\cite{Bohr98}.
For instance, precision mass measurements of exotic nuclei have led to discoveries of the disappearance of the neutron magic number at  $N=20$~\cite{Thi75} and the rise of the new sub-shell closure at $N=32$~\cite{Wien13,N=32_Sc}.
The masses of extremely exotic nuclei are used to determine the borders of nuclear existence, the drip-lines \cite{Novikov2002,Neufcourt},
as well as new mass measurements provide valuable benchmarks for nuclear theories \cite{Sobiczewski2018}.
In nuclear astrophysics the needed ground state properties of many nuclides involved in
the rapid neutron capture (r-process) or the rapid proton capture (rp-process) processes still have to be measured \cite{Schatz2013}.
In $\beta$-decay experiments, the Fermi (F) and Gamow-Teller (GT) transition strengths are deduced from the measured $\beta$ feedings
as well as the decay $Q$-values~\cite{GT_cal}.
The latter are determined via the mass differences of the corresponding parent and daughter nuclei.
Accurate nuclear masses in the lighter $Z=N$ region are often used to test the validity of the isobaric multiplet mass equation (IMME)~\cite{2014MA56,2013Lam},
which is associated with isospin symmetry in particle and nuclear physics.
If a breakdown of the IMME is found, this may offer a possibility to study mechanisms responsible for the isospin-symmetry breaking~\cite{Bentley07}.

Exotic nuclei of interest today are typically short-lived and have tiny production rates.
Therefore mass measurements of such short-lived and rare nuclei inevitably require very sensitive and fast experimental techniques.
One of such techniques is isochronous mass spectrometry (IMS) applied to nuclei stored in a heavy-ion storage ring~\cite{IMS}.
In the past few years, the masses of a series of $T_{z}=-1$ and $-3/2$ short-lived proton-rich nuclei in the $fp$ shell have been measured by employing IMS at the Cooler Storage Ring (CSRe) in Lanzhou~\cite{Review_Zhang,Regular_article_Zhang,2012ZH34}.
As a continuation of this work, we report here precision mass measurements of $T_{z}=-2$ $fp$-shell nuclei
produced in the projectile fragmentation of $^{58}$Ni.
The paper is organized as follows: Experimental details and data analysis are described in Sec.~\ref{Experiment and Data Analysis}. The new results and their impact on nuclear structure are given in Sec.~\ref{Experimental results} and Sec.~\ref{discussion}, respectively. Summary and conclusions are given in Sec.~\ref{summary}.

\section{Experiment and Data Analysis}
\label{Experiment and Data Analysis}

The experiment was performed at the HIRFL-CSR acceleration complex~\cite{Xiajiawen},
which consists of a separated sector cyclotron (SSC, $K=450$),
a sector-focusing cyclotron (SFC, $K=69$),
a main cooler-storage ring (CSRm) operating as a heavy-ion synchrotron,
and an experimental cooler-storage ring CSRe.
The two storage rings are connected together by an in-flight fragment separator RIBLL2.
The CSRm has a circumference of 161.00 m and a maximum magnetic rigidity $B\rho=12.05$ Tm.
Hence, $^{12}$C$^{6+}$ and $^{238}$U$^{72+}$ ions can be accelerated to energies of about 1 GeV/u and 500 MeV/u, respectively.
The CSRe has a circumference of 128.80 m and a maximal magnetic rigidity $B\rho=9.40$ Tm~\cite{WenLongZHAN}.

In this experiment, 468~MeV/u $^{58}$Ni$^{19+}$ primary beams of about $8\times 10^7$ particles per spill
were fast-extracted from the CSRm and were fragmented in a $\sim$15~mm thick $^9$Be target placed in front of the RIBLL2.
The reaction products from the projectile fragmentation of $^{58}$Ni emerged the target with relativistic energies mostly as bare nuclei.
They were selected and analysed in flight with the RIBLL2.
The CSRe was tuned into the isochronous ion-optical mode~\cite{Haus00,Tuxiaolin} with the transition energy $\gamma_t=1.400$.
The CSRe was set to a fixed magnetic rigidity of $B\rho=5.5778$~Tm such that $^{44}$Cr$^{24+}$ ions
fulfill the isochronous condition of $\gamma=\gamma_t$, where $\gamma$ is the relativistic Lorenz factor.
The required energy of the primary beams and the magnetic rigidity of the RIBLL2 were determined with LISE++ simulations~\cite{Tar08}
to achieve the maximum yield and the optimal transmission for $^{44}$Cr$^{24+}$ ions.
Every 25 s, a fresh primary beam was fast extracted to produce the nuclides of interest. Only a small fraction of the projectile fragments lying within the $B\rho$ acceptance of $\pm 0.2$\% of the RIBLL2-CSRe system
were transmitted and stored in the CSRe. In each injection, up to 18 ions were selected and injected into the CSRe for a measurement of 300 $\mu$s.

\begin{figure}[htbp]
  \centering
	\includegraphics[width=0.8\columnwidth]{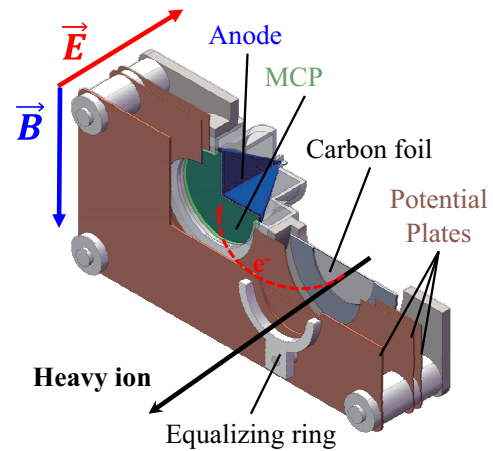}
	\caption{(Color online). Schematic view of the time-of-flight detector~\cite{Meibo} installed inside the CSRe aperture.
    The electric field is generated by the potential plates and an equalising ring.
    The magnetic field is produced by Helmholtz coils (not shown) placed outside the ultra-high vacuum environment of the CSRe.
 }
  \label{detector}
\end{figure}

A time-of-flight (TOF) detector~\cite{Meibo} was installed inside the CSRe aperture to measure the revolution times of stored ions.
The detector was equipped with a 19 $\mu$g/cm$^2$ carbon foil of 40 mm in diameter
and a micro-channel plate (MCP) with a fast timing anode (see Fig.~\ref{detector}).
Passages of swift ions through the carbon foil caused secondary electrons released from the foil surface.
The number of such secondary electrons depends on the electronic stopping power of the passing ion~\cite{Rothad}, \emph{dE/dx},
which at relativistic energies is roughly proportional to the square of its atomic number $Z$.
Combined with the geometrical efficiency of the MCP,
the overall detection efficiency of the TOF detector for a single ion passage varied from 7$\%$ to 80$\%$ depending on the ion species.
Secondary electrons were isochronously guided to the MCP by perpendicularly arranged electric and magnetic fields.
The timing signals from the anode were directly recorded
by a high-performance digital oscilloscope Agilent DSO90604A (20 GS/s rate, 6 GHz analog bandwidth).
Typical fall-times of the negative-voltage timing signals were $250-500$ ps.
The time stamps were extracted by using the CFD (Constant Fraction Discrimination) technique applied to the digitised timing signals.
Finally, the time sequences, i.e. the passage times as a function of the revolution number, were obtained for each individual ion.
Only the time sequences containing more than 30 timestamps within a circulation time of more than 100 $\mu$s were considered in the data analysis.
The time sequences were fitted with a second order polynomial function.
The revolution times were obtained as a slope of the fit curve at the 35th revolution.
More details of the typical signal processing and data analysis can be found in Ref.~\cite{Tuxiaolin}.
Since the magnetic fields of the CSRe magnets slowly drifted during the experiment,
the field-drift correction procedure developed in Refs.~\cite{Regular_article_Zhang,magnetic Cor} has been implemented.
The corrected revolution times were put into a histogram forming a revolution-time spectrum.

\begin{figure*}[htbp]
  \centering
	\includegraphics[width=2.0\columnwidth]{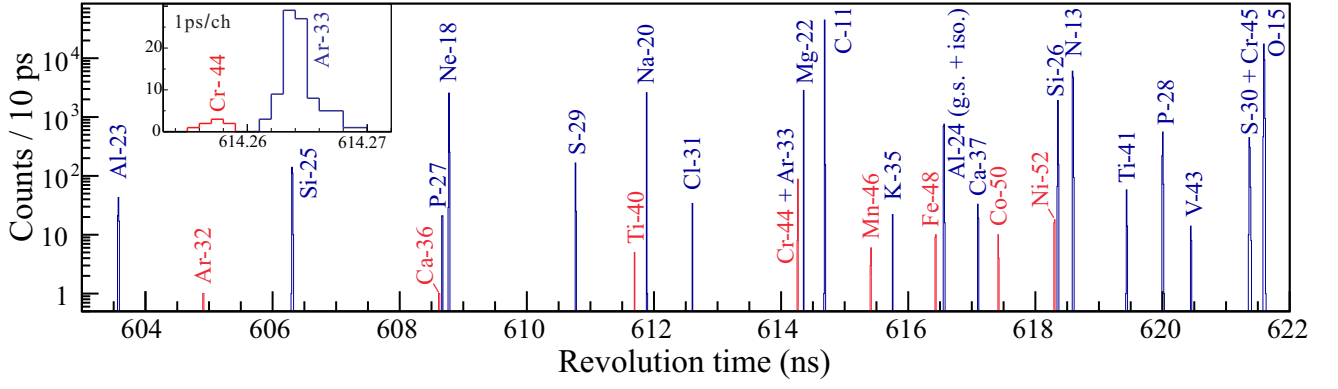}
	\caption{(Color online). A part of the measured revolution-time spectrum.
    $T_{z}=-2$ nuclei are indicated with red color.
    The inset shows an expanded time range around $^{44}$Cr and $^{33}$Ar illustrating the corresponding revolution-time peaks well resolved.}
  \label{spectrum}
\end{figure*}
\begin{figure}[htbp]
  \flushleft
	\includegraphics[width=1.0\columnwidth]{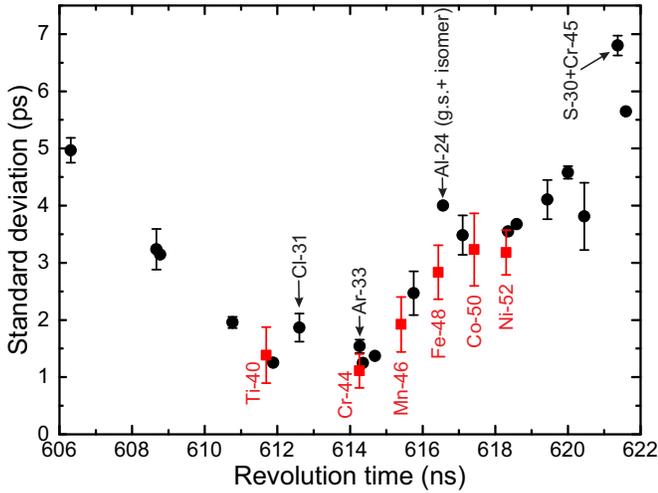}
	\caption{(Color online). Obtained standard deviations (RMS) of the
revolution-time peaks shown in Fig.~\ref{spectrum}. Red squares are the values for $T_{z}=-2$ nuclides.
The label Al-24 (g.s.+isomer) indicates that the corresponding peak represents a mixture of the ground- and a
low-lying isomeric state. The label S-30+Cr-45 corresponds to the RMS value deduced from the peak of unresolved $^{30}$S and $^{45}$Cr nuclei.}
  \label{sigma_T}
\end{figure}
A part of the corrected revolution-time spectrum in a time window of 603 ns $\le t \le$ 622 ns is shown in Fig.~\ref{spectrum}.
The particle identification has been done according to the procedures described in Refs.~\cite{Tuxiaolin,magnetic Cor}.
The inset of Fig.~\ref{spectrum} shows the zoomed time range around $^{44}$Cr and $^{33}$Ar,
where one sees that $^{44}$Cr and $^{33}$Ar with nearly the same $m/q$ values (${\triangle (m/q)}/{(m/q)}\approx 2\times 10^{-5}$) are well separated.
The corresponding standard deviations (or equivalently the root mean squares, RMS) of each peak are shown in Fig.~\ref{sigma_T} and range between 1 and 5 ps.
The parabolic shape of the RMS values versus revolution times is well-understood.
The minimum RMS value is found around $^{44}$Cr, for which the isochronous tuning of the CSRe was done.
The widths of the revolution-time peaks increase when moving away from $^{44}$Cr.
\begin{figure}[htbp]
	\flushleft
	\includegraphics[width=1.0\columnwidth]{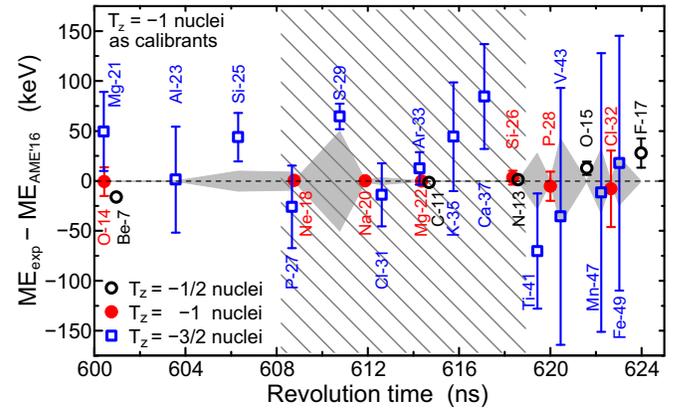}	
	\caption{(Color online). Experimental $ME$ values determined in this work compared with the literature values from the latest Atomic-Mass Evaluation AME$^{\prime}$16~\cite{AME2016} and Ref.~\cite{27S} for $^{27}$P. The $T_z=-1$ nuclides (red filled circles) were used as calibrants. The shaded area indicates the time range where no systematic deviations were observed (see text). The grey shadowed areas indicate the 1$\sigma$ mass uncertainty in AME$^{\prime}$16 and in Ref.~\cite{27S} for $^{27}$P.
	}
	\label{systematic_deviation}
\end{figure}

Four series of nuclides with $-2\leq T_{z}\leq-1/2$ are shown in Fig.~\ref{spectrum}.
Most of the observed nuclides have well-known masses except for $T_{z}=-2$ nuclides.
To estimate possible systematic uncertainties, $T_z=-1$ nuclides were used to calibrate the time spectrum via the expression
\begin{eqnarray}\label{calibra equ}
m/q(t)=a_{0}+a_{1}\cdot t+a_{2}\cdot t^{2}+a_{3}\cdot t^{3},
\end{eqnarray}
where $a_{0}$, $a_{1}$, $a_{2}$ and $a_{3}$ are free parameters.
The well-known masses of $T_z=-1/2,-3/2$ nuclides were re-determined and compared with the literature values~\cite{AME2016} in Fig.~\ref{systematic_deviation}.
This comparison revealed, that within the fitting range of 608 ns $\leq t \leq $ 619 ns for $T_z=-1/2,-3/2$ nuclides
no additional systematic uncertainty is required (see methodology in Sec.~\ref{Experimental results}).
However, beyond $t\sim 619$ ns a systematic deviation is observed for $T_z=-1/2$ nuclides.
Such deviation is not evident for $T_z=-3/2$ nuclides at the level of the present experimental uncertainties. The mass of $^{29}$S compiled in AME$^{\prime}$16~\cite{AME2016} is from the earlier work using $^{32}\rm S(^3\rm He,^6\rm He)^{29}\rm S$ reaction~\cite{Ben73}. The deviation to the obtained mass value of $^{29}$S in this experiment has been reported and discussed in Ref.~\cite{FuCY}.

In the final calibration all nuclides with known masses in the time window 608 ns $\leq t \leq $ 619 ns have been used except for $^{27}$P and $^{29}$S.
The masses of $T_{z}=-2$ $fp$-shell nuclei, i.e. $^{40}$Ti, $^{44}$Cr, $^{46}$Mn, $^{48}$Fe, $^{50}$Co and $^{52}$Ni, were determined and converted~\cite{Regular_article_Zhang,magnetic Cor} into atomic mass excesses defined as $ME=(m-Au)c^2$.
\section{Experimental results}\label{Experimental results}
\begin{table*}[htbp]
	\caption{
		Experimental $ME$-values obtained in this work and from the AME$^{\prime}$16~\cite{AME2016}.
		The number of ions identified, $N$, and the standard deviations of the time peaks in Fig.~\ref{spectrum}, $\sigma_{t}$, are listed in the second and third columns, respectively.
		The deviations, $\delta ME= ME_{CSRe}-ME_{AME^{\prime}16}$, are given in the sixth column.
		The predicted $ME$s from the IMME are given in the last column.
		}
	\begin{ruledtabular}
		\begin{tabular}{lrrrrrr}
			Atom      & $N$  &$\sigma_{t}$& $ME\rm_{CSRe}$ & $ME\rm_{AME^{\prime}16}$~~ & $\delta ME$~~ &  IMME \\
			&      &(ps)        & (keV)       & (keV)~~                  & (keV)~~  &    (keV)           \\
			\hline
			$^{40}$Ti & 5    & 1.39       & $-9025(75)$   & $-8850(160)^{~~}$            & $-175(177)^{~~}$ & $ -9060(6)$  \\
			$^{44}$Cr & 8    & 1.11       & $-13422(51)$   &$-13360(300)$\footnote{Extrapolated values in AME$^{\prime}$16~\cite{AME2016}.}  &  $-62(304)$\footnotemark[1]    &  $ -13484(19)$     \\
			$^{46}$Mn & 9    & 1.92       &$-12418(87) $  & $-12570(400)$\footnotemark[1]   & 152(409)\footnotemark[1]  &$-12493(30)$     \\
			$^{48}$Fe & 19   & 2.83       & $-18009(92) $  &$-18000(400)$\footnotemark[1]   &$-9(410)$\footnotemark[1]    & $-18097(14) $   \\
			$^{50}$Co & 14   & 3.23       & $-17589(126)$  &$-17630(400)$\footnotemark[1]   &41(419)\footnotemark[1]    & $-17552(24)$   \\
			$^{52}$Ni & 34   & 3.18       & $-22560(83)$   &$-22330(400)$\footnotemark[1]   &$-230(409)$\footnotemark[2]  & $-22699(22) $    \\
		\end{tabular}
	\end{ruledtabular}
	\label{Mass table}
\end{table*}

\begin{figure}[htbp]
	\flushleft
	\includegraphics[width=1.0\columnwidth]{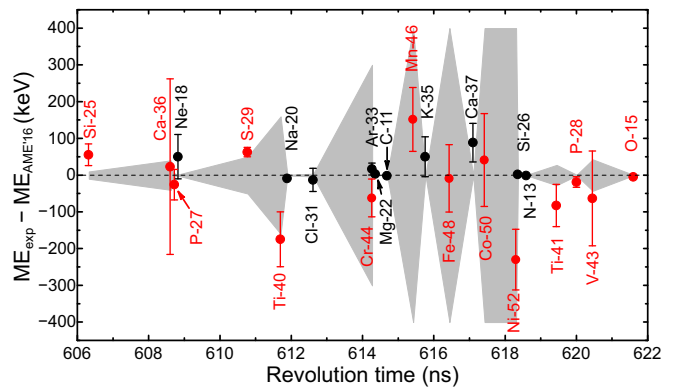}
	\caption{(Color online). Differences between experimental $ME$-values determined in this work and those from the Atomic-Mass Evaluation AME$^{\prime}$16~\cite{AME2016} and Ref.~\cite{27S} for $^{27}$P.
	Mass values for each of the ten reference nuclides (filled black circles) were re-determined by using the other nine nuclides as calibrants.
	The red circles correspond to the newly determined masses when all ten reference nuclides were employed for mass calibration.
	The grey shadowed areas indicate the 1$\sigma$ mass uncertainty in AME$^{\prime}$16 and in Ref.~\cite{27S} for $^{27}$P.
	}
	\label{calibration}
\end{figure}

Newly determined masses of $T_{z}=-2$ $fp$-shell nuclei are given in Table~\ref{Mass table}
together with their literature values from AME$^{\prime}$16~\cite{AME2016} and Ref.~\cite{27S} for $^{27}$P.
The new and re-determined masses are compared with the literature values in Fig.~\ref{calibration}.
The black filled symbols in this figure indicate the nuclei used for the calibration.
Each of the $N_{c}=10$ $ME$-values of the reference nuclides was re-determined by using the other nine nuclides as calibrants.
This technique is referred to as leave-one-out cross-validation method~\cite{Tuxiaolin}.
The normalised $\chi_{n}$ defined as
\begin{eqnarray}\label{Chi-square equ}	\chi_{n}=\sqrt{\frac{1}{N_c}\sum\limits_{i=1}^{N_c}\frac{[(\frac{m}{q})_{i,exp}-(\frac{m}{q})_{i,AME}]^{2}}{[\sigma_{exp}(\frac{m}{q})_{i}]^{2}+[\sigma_{AME}(\frac{m}{q})_{i}]^{2}}}~~,
\end{eqnarray}
was found to be $\chi$$_{n}=1.14$.
This value is within the expected range of $\chi_{n}=1\pm1/\sqrt{2N_{c}}=1\pm0.22$ at $1\sigma$ confidence level,
thus indicating that no additional systematic uncertainty needs to be considered.

To show the accuracy/reliability of the present results, the well-known $ME$-values for nuclides lying outside the considered fitting range, namely  $^{32}$Ar, $^{25}$Si, $^{36}$Ca, $^{27}$P~\cite{27S}, $^{41}$Ti, $^{28}$P, $^{43}$V, and $^{15}$O, were as well re-determined by extrapolating the fit function. The obtained results are given in Fig.~\ref{calibration} and show good agreement with the literature values~\cite{AME2016,27S}. Due to a long-range extrapolation, our mass of $^{32}$Ar agrees only within $3\sigma$ with the result in Ref.~\cite{33Ar}.

The mass of $^{40}$Ti was previously measured in $^{40}$Ca($\pi^{+}$,$\pi^{-}$)$^{40}$Ti double-charge-exchange reactions~\cite{40Ti}, and the experimental result of $ME(^{40}$Ti)$=-8850(160)$ keV has been adopted in the AME$^{\prime}$16~\cite{AME2016}.
Our measurement yields $ME(^{40}$Ti)$=-9025(75)$ keV which is in agreement with the adopted value,
though the precision is improved by a factor of 2.1.

The masses of $^{44}$Cr, $^{46}$Mn, $^{48}$Fe, $^{50}$Co and $^{52}$Ni were measured for the first time in this work.
For the even-even nuclei, the $ME$-values given in Table~\ref{Mass table} correspond to the ground states.
It is worth noting that the revolution times of $^{44}$Cr and $^{33}$Ar are very close to each other (see the inset in Fig.~\ref{spectrum}).
The mass of $^{33}$Ar was re-determined to be $ ME(^{33}{\rm Ar})=-9368(16)$~keV which is in excellent agreement with the literature value~\cite{33Ar}.
This provides an additional evidence for the reliability of the present measurement of $^{44}$Cr.

\begin{figure}[htbp]
	\centering
	\includegraphics[width=0.8\columnwidth]{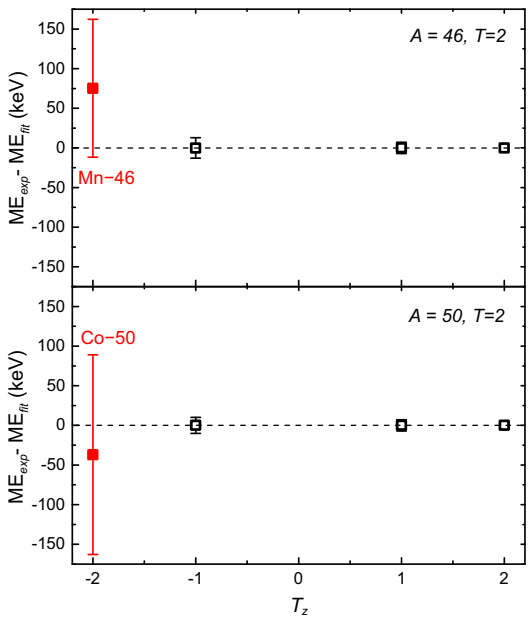}
	\caption{(Color online). Comparison of the experimental $ME$-values of $^{46}$Mn and $^{50}$Co with the IMME predictions. }
	\label{A=46 and 50}
\end{figure}
In the case of the odd-odd nuclei $^{46}$Mn and $^{50}$Co, the contamination by low-lying isomeric states cannot be excluded.
An isomeric state with excitation energy $E_{x}=142.528(7)$ keV and half-life $T_{1/2}=18.75(4)$~s \cite{Nubase2016} is known in $^{46}$Sc,
which is the mirror nucleus of $^{46}$Mn.
Taking into account the mirror symmetry, a low-lying isomer at $E_{x}\sim 150$ keV may exist in $^{46}$Mn.
Although low-lying isomers have been observed in odd-odd $^{52}$Co and $^{54}$Co,
no such isomers were observed in mirror nuclei $^{50}$Co and $^{50}$V.
In the experimental revolution-time spectrum, single peaks without obvious broadenings were observed for $^{46}$Mn and $^{50}$Co.
Furthermore, the extracted peak widths follow the expected systematic behaviour (see Fig.~\ref{sigma_T}), though the counting statistics is low.

The masses of $^{46}$Mn and $^{50}$Co could be calculated by using the isospin multiplet mass equation (IMME)~\cite{Weinberg,Wigner57} expressed as
\begin{eqnarray}\label{IMME equ}
ME(\alpha,T,T_{z}) = a(\alpha,T)+b(\alpha,T)\cdot T_{z}+c(\alpha,T)\cdot T_{z}^{2},
\end{eqnarray}
where $ME$s are mass excesses of isobaric analog states (IAS) of a multiplet with fixed mass number $A$ and total isospin $T$.
The coefficients \emph{a, b, c} depend on $A$, $T$ and other quantum numbers such as the spin-and-parity $J^\pi$,
but are independent from $T_z$.
This mass equation is considered to be accurate within mass uncertainties of a few tens of keV.
The $ME$-values of three IAS~\cite{Nubase2016} were used to determine the \emph{a, b, c} coefficients.
The calculated $ME$-values of $^{46}$Mn and $^{50}$Co are given in the last column of Table~\ref{Mass table}
and are compared to the experimental results in Fig.~\ref{A=46 and 50}.
Excellent agreement within one standard deviation is obtained.
It is therefore suggested that the measured here $ME$-values correspond to the ground states.
An additional argument supporting this suggestion is that any contamination by an unknown isomeric state would
lead to lower mass excess values than those given in Table~\ref{Mass table}.
\section{Discussion}\label{discussion}
\subsection{Test of nuclear mass models}
\begin{figure}[htbp]
  \centering
	\includegraphics[width=0.8\columnwidth]{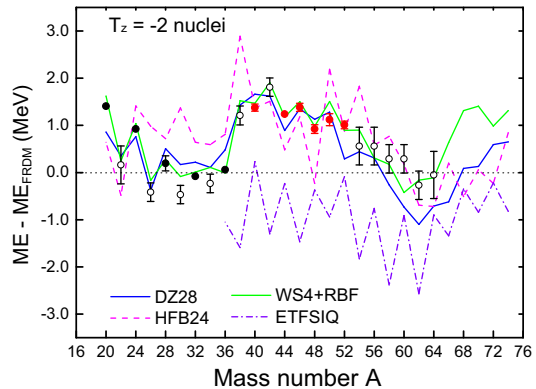}
	\caption{(Color online). Comparison of the new experimental mass values with predictions of several mass models for $T_{z}=-2$ nuclei.
	The adopted and extrapolated mass values from AME$^{\prime}$16~\cite{AME2016} are presented with the filled and open black circles, respectively.
	The filled red circles indicate the new experimental masses of this work. }
  \label{global mass models}
\end{figure}

The nuclear masses measured in this work can be used to test modern nuclear mass models.
The accuracy of current theoretical models has been recently investigated in Refs.~\cite{Sobiczewski2018,Sobi14}.
Among the ten often-used models of various nature, the macroscopic-microscopic models of Wang and Liu~\cite{Liu11,Wang11}
and of Duflo and Zuker (DZ28)~\cite{DZ28} were found to be the most accurate in various mass regions
characterised by the smallest RMS values of $250\sim 500$~keV.
Figure~\ref{global mass models} shows a comparison of the new experimental masses of $T_z=-2$ nuclei
with the predictions of five global mass models, namely
the finite-range droplet model (FRDM) of M\"{o}ller and Nix ~\cite{FRDM, FRDM2},
the Duflo and Zuker (DZ28) mass model,
the Hartree-Fock-Bogoliubov calculations with Skyrme interaction BSk24 (HFB-24)~\cite{HFB},
the Extended-Thomas-Fermi-Strutinski-Integral mass table with introduced quenching of shell closures (ETFSI-Q)~\cite{ETFSIQ},
and the latest version of the model of Wang and Liu labeled as WS4 with the radial basis function correction~\cite{Wang14}.

Inspection of Fig.~\ref{global mass models} leads to the same conclusion as in Refs.~\cite{Sobiczewski2018, Sobi14}
that the DZ28 and WS4 mass models give the most accurate mass predictions for $T_z=-2$ nuclei.
Especially, the zig-zag staggering in the 40 $\le A \le$ 52 region ($pf$-shell) can be well described
by the WS4 calculations, confirming the high predictive power of the model.

In contrary to the global mass models aiming at describing the entire mass surface,
local mass relations are often more accurate in near extrapolations into unknown masses.
Such relations are, for example, the IMME~\cite{Weinberg,Wigner57},
the Audi-Wapstra systematics~\cite{Audi93},
the Garvey-Kelson (G-K) mass relations~\cite{GK,GK69,Barea08,Bao13,Cheng14a},
the mass relations based on the residual proton-neutron interactions~\cite{Fu11,Jiang12,Cheng14b},
and the mass formulas connecting mirror nuclei based on the isospin conservation~\cite{Bao16, Zong19}.
An improved approach of Refs.~\cite{Bao16, Zong19} gives a RMS value as small as $93$ keV~\cite{Zong20}.

\begin{figure}[htbp]
  \centering
	\includegraphics[width=0.8\columnwidth]{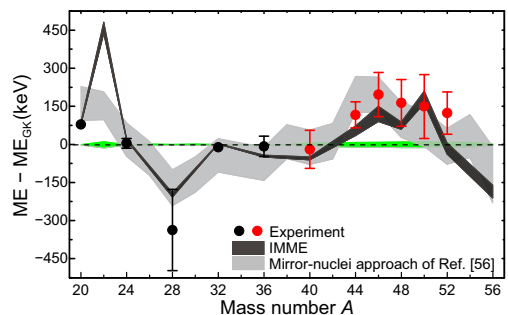}
	\caption{(Color online). Comparison of the new experimental $ME$-values of $T_{z}=-2$ nuclei with predictions by local mass relations using the mass values from AME$^{\prime}$16~\cite{AME2016,Nubase2016}.
	The literature masses~\cite{AME2016,Nubase2016} are shown with black filled circles.
	The new masses from Table~\ref{Mass table} are indicated with red symbols.
	The shaded areas represent 1$\sigma$ uncertainty of the theoretical predictions.
 }
  \label{local models graph}
\end{figure}
Taking the Garvey-Kelson predictions as reference, Figure~\ref{local models graph} shows the comparison of the new experimental results
with predictions by the IMME and the G-K relations as well as the mirror-nuclei approach~\cite{Zong20}.
The predictions were obtained by using the known-mass values from AME$^{\prime}$16~\cite{AME2016}.
The newly measured $ME$-values are in good agreement with the mass predictions from both the IMME and the mirror-nuclei approach~\cite{Zong20}, but are systematically higher than those from G-K relations except for $^{40}$Ti.
\subsection{Validity of the isospin multiplet mass equation}
Although the quadratic form of the IMME, i.e. Eq.~(3), is commonly considered to be accurate,
precision mass measurements can be used for testing its validity~\cite{Review_Zhang}.
Typically one adds to Eq.~(3) extra terms such as $d \cdot T_z^3$ or/and $e\cdot T_z^4$,
which provide a measure of the breakdown of the quadratic form of the IMME.
By taking into account the second-order Coulomb effects, 3-body interactions, and isospin mixing,
the $d$ and $e$ coefficients have generally been expected to be smaller than a few keV~\cite{Henley69,Janecke69,Bertsch70,Dong18,Dong19}.
Numerous measurements have been performed investigating the validity of the IMME.
Reviews and compilations of existing data can be found in Refs.~\cite{2014MA56,2013Lam} and references cited therein.
For $T=2$ isospin quintets, $A=40$ was the heaviest multiplet with all masses experimentally known.

\begin{figure}[htbp]
  \centering
	\includegraphics[width=0.8\columnwidth]{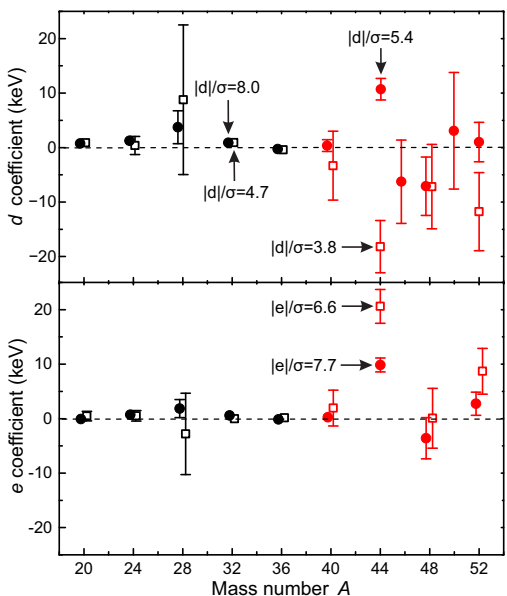}
	\caption{(Color online). $d$- and $e$-coefficients obtained from the least-squares fitting using the quartic (empty squares) form of the IMME.
	Results of 4-parameter fitting with only $d$- or $e$-coefficient included are shown with filled symbols.
	$ME$-values from the AME$^{\prime}$16 (black symbols)~\cite{AME2016,Nubase2016} and from this work (red symbols) were used.
	For the $A=36$ quintet the mass value of $^{36}$Ca was taken from Ref.~\cite{Ca36}.
 }
  \label{d coefficients}
\end{figure}

By adding the new nuclear masses obtained in this work, experimental information for $A=40, 44, 48,$ and $52$ $T=2$ quintets is now completed,
which includes the heaviest quintet in the $pf$ shell.
The data were used to extract $d$- and $e$-coefficients which are shown in Fig.~\ref{d coefficients} as empty squares.

For $A=46,50$ quintets, only four-parameters fitting with either $d$- or $e$-coefficient included was possible.
The corresponding results for $d$- and $e$-coefficients are plotted in Fig.~\ref{d coefficients} with filled symbols.
Except for $A=44$, all obtained coefficients are compatible with zero at 2$\sigma$ confidence level.
For the $A=44$ quintet, the results are $|d|/\sigma_d=3.8$ and $|e|/\sigma_e=6.6$.
Although a large $|d|/\sigma_d$ is observed for the $A=32$ quintet,
the magnitude of the $d$-coefficient is small and can theoretically be reproduced by taking into account the isospin mixing~\cite{A=32,2013LA15}.
Large $d$- and $e$-coefficients for the $A=44$ quintet, which significantly deviate from zero, are surprising.
Since the $d$-coefficient is modified from $-18.2(4.8)$ for the 5-parameter fit to
$+10.7(2.0)$ for the 4-parameter fit, the convergence of the fitting procedure is questionable.

\subsection{$\beta$-delayed proton decays of $^{44}$Cr and the isobaric analog state in $^{44}$V}\label{44Cr}
A large $d$-coefficient has already been obtained by using the $T=2$, IAS in $^{44}$V assigned
via the observations of $\beta^+$-delayed protons from $^{44}$Cr decay~\cite{Dossat}.
However, to restore the validity of the IMME, the authors of Ref.~\cite{Dossat} suggested
that the state in $^{44}$Ti at $E_x=9298$ keV, rather than at $E_x=9388$ keV, is the $T=2$ IAS of the ground state of $^{44}$Cr.
If the suggested value is used to extract $a,b,c $ coefficients in the quadratic form of the IMME,
the calculated mass excess of $^{44}$Cr is $-13620(20)$ keV which is 198(51) keV ($\sim 4 \sigma$)
smaller than our new experimental value of $-13422(51)$ keV.
Furthermore, if the 5-parameters fitting is performed,
the corresponding $d$- and $e$-coefficients significantly deviate from zero ($d=-18.2(4.8)$, $e=10.6(3.1)$).
These results contradict theoretical expectations in the framework of isospin symmetry~\cite{Henley69,Janecke69,Bertsch70,Dong18,Dong19}.

The masses of $^{44}$Ca, $^{44}$Sc, and $^{44}$Ti are known with high precision~\cite{AME2016}.
The $T=2$ IAS in $^{44}$Sc was identified in several experiments and is located at $E_x=2778(3)$ keV~\cite{ENSDF}.
This value has been confirmed in a recent investigation through a $(p,n)$-type $^{44}$Ca($^3$He,t)$^{44}$Sc reaction~\cite{Fujita13}.
The first $T=2$ IAS in the $Z=N$ self-conjugate nucleus $^{44}$Ti was identified via the isospin-allowed $^{46}$Ti$(p,t)^{44}$Ti reaction~\cite{Garvey64,Bapaport72}.
It was later assigned to $E_x=9338(2)$ keV through $\gamma$-decay spectroscopy in the $^{40}$Ca$(\alpha,\gamma)^{44}$Ti reaction~\cite{Simpson72}.
Further studies have led to the identification of a close-lying $J^{\pi}=0^+$ state below the $T=2$ IAS in $^{44}$Ti~\cite{Moalem76}.
This state has been placed at $E_x=9298(2)$ keV through $\gamma$-decay measurements~\cite{Dixon78},
thus forming an isospin-mixed doublet.
Decay-width analyses including the $\gamma$-decay branching ratios revealed that the main $T=2$ strength remains in the 9338(2) keV level.
This state was interpreted to originate from the mixing of the unperturbed $T=2$, IAS at $E_x=9330(4)$ keV
with a state at $E_x=9306(4)$ keV with $T=1$ or $T=0$~\cite{Dixon78}.

By using the $ME$-values of $T=2$ IAS in $^{44}$Ca, $^{44}$Sc, and $^{44}$Ti mentioned above and the quadratic form of the IMME,
the $ME$-values of $T=2$ IAS in $^{44}$Cr and $^{44}$V were calculated to be $-13412(31)$ keV and $-21010(14)$ keV, respectively.
The former value is in excellent agreement with $-13422(51)$ keV obtained in this work, whereas the latter value is 114 keV ($8.8\sigma$)
larger than the value given in Ref.~\cite{Dossat}.
\begin{table*}[htbp]
	\caption{
    Compilation of $\beta$-delayed proton decay energies, $Q_p$, $\gamma$-ray energies, $E_{\gamma}$, as well as their branching ratios, $I_p$ and $I_{\gamma}$, for the decay of $^{44}$Cr. The results from Refs.~\cite{Dossat,TPC} are converted to the centre-of-mass energies.
    The weighted-average values are adopted in the table.
	}
	\begin{tabular*}{18cm}{@{\extracolsep{\fill}}lrrcrrcrrcrrrrrr}
		\hline
		&\multicolumn{2}{c}{Ref.~\cite{Dossat}}&&\multicolumn{2}{c}{Ref.~\cite{TPC}}&&\multicolumn{2}{c}{ Weighted Average}&& \multicolumn{6}{c}{Proposed Assignment} \\
\cline{2-3}\cline{5-6}\cline{8-9}\cline{11-16}
		&$Q_{p}$~(keV) &$I_{p}$~(\%)&&$Q_{p}$~(keV) &$I_{p}$~(\%)&&$Q_{p}$~(keV) &$I_{p}$~(\%)&&\multicolumn{3}{l}{This work}&\multicolumn{3}{l}{Ref.~\cite{Dossat}} \\
	 \hline
	   1&              &            && 759(26)      & 0.6(2)     && 759(26)      & 0.6(2) && \multicolumn{3}{l}{$1^+_2$ to $(3/2^-)$ in $^{43}$Ti}&\multicolumn{3}{l}{}\\
	    &              &            &&              &            &&              &        && \multicolumn{3}{l}{or IAS to $(3/2^+)$ in $^{43}$Ti}&\multicolumn{3}{l}{}\\
	   2& 910(11)      & 1.7(3)     && 917(53)      & 2.7(5)     && 910(11)      & 2.0(3) && \multicolumn{3}{l}{$1^+_3$ to $(3/2^-)$ in $^{43}$Ti}&\multicolumn{3}{l}{IAS to g.s. in $^{43}$Ti}\\
	   3& 1384(12)     & 1.1(3)     && 1371(62)     & 1.4(3)     && 1384(12)     & 1.3(3) && \multicolumn{3}{l}{$1^+_4$ to $(3/2^-)$ in $^{43}$Ti}&\multicolumn{3}{l}{} \\
	   4& 1741(15)     & 0.6(3)     && 1719(44)     & 0.5(2)     && 1739(14)     & 0.5(2) && \multicolumn{3}{l}{$1^+_5$ to $(3/2^+)$ in $^{43}$Ti }&\multicolumn{3}{l}{} \\
		\hline
        &$E_{\gamma}$~(keV)&$I_{\gamma}$~(\%)&&$E_{\gamma}$~(keV)&$I_{\gamma}$~(\%)&&$E_{\gamma}$~(keV)&$I_{\gamma}$~(\%)&&&& \\
        \hline
	   1& 676.9(3)     & 59(5)      &&              &            && 676.9(3)     &   59(5)&&   \multicolumn{3}{l}{}&\multicolumn{3}{l}{$^{44}$V:$1^{+}_1$ to 2$^{+}$(g.s.)}\\
		\hline
	\end{tabular*}
	\label{summary of proton and gamma}
\end{table*}

\begin{figure*}[htbp]
	\includegraphics[width=2.0\columnwidth]{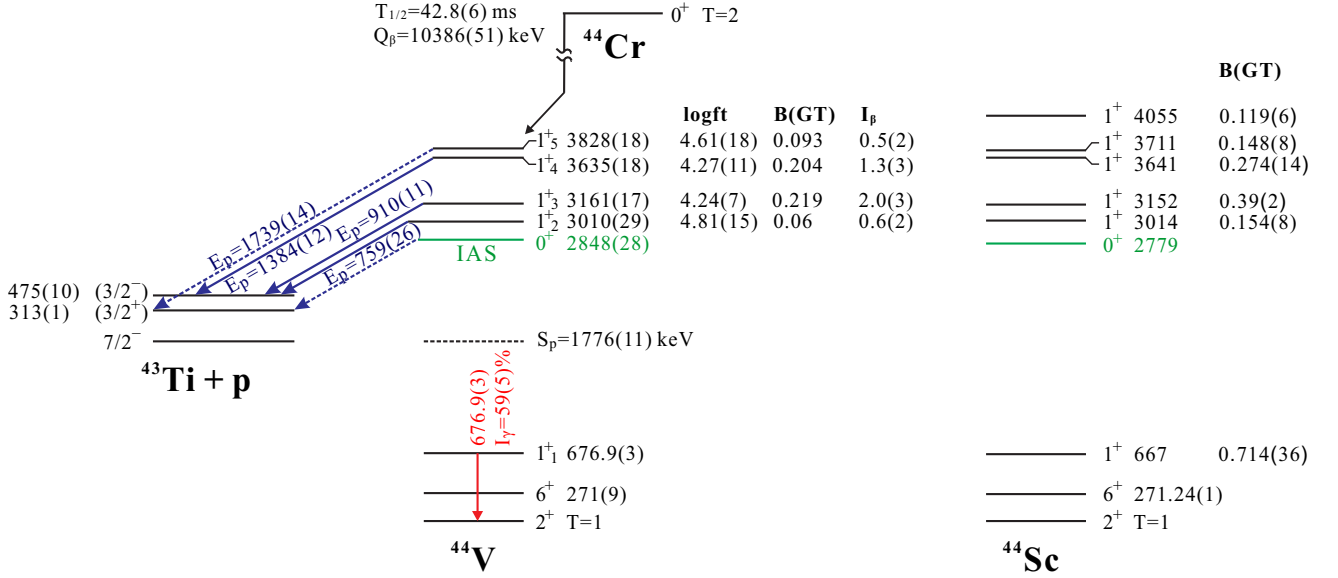}
	\caption{(Color online). (Left) Proposed partial decay scheme of $^{44}$Cr. (Right) The level structure of $^{44}$Sc, the mirror nucleus of $^{44}$V, identified in the $^{44}$Ca($^{3}$He,$t$)$^{44}$Sc reaction~\cite{Fujita13}. Only levels with B(GT) $\geq$ 0.1 are shown. The $\beta$ feedings $(I_{\beta})$ and log $ft$ values of the $^{44}$Cr $\beta$-decays are deduced from the branching ratios of $\beta$-delayed protons and $Q$-values suggested in this work. The $\beta$-decay B(GT) values are calculated from Eq.~(\ref{B(GT)}). The levels shown in green colour indicate the excited $T=2,~J^{\pi}=0^{+}$ IAS. All energies are in keV.}
	\label{decay scheme}
\end{figure*}
\begin{table}[htbp]
	\caption{
		Compilation of the $ME$-values for the ground states of $^{44}$Cr and $^{44}$Ca, and for the lowest $J^\pi=0^+$, $T=2$ IAS in $^{44}$V, $^{44}$Ti, and $^{44}$Sc. The data are taken from AME$^{\prime}$16 and NUBASE$^\prime$16 \cite{AME2016,Nubase2016} except for $^{44}$Cr, $^{44}$V, and $^{44}$Ti. The unperturbed level energy of the IAS in $^{44}$Ti~\cite{Dixon78} is used. The weighted-average mass of the ground state of $^{44}$V measured in Refs.~\cite{Regular_article_Zhang,LEBIT_44V} is adopted.  
		The parameters of the IMME fits are listed in Table~\ref{IMME_fit}.
	}
	\begin{tabular*}{8.65cm}{lcrrr}
		\hline
		~Atom~~~~~ &~~~$T_{z}$~~~& ~~~~~~~~~~~~$ME$(g.s.)~& ~~~~~~~~~~~~~~~~~~$E_{x}$~~& ~~~~~~~~~~$ME$(IAS)~~\\
		&             & (keV)~                  & (keV)~                       & (keV)~ \\
		\hline
		$^{44}$Cr  & $-2$        & $-13422(51)^{*}$       & $0^{~~}$                    & $-13422(51)^{*~}$ \\
		$^{44}$V   & $-1$        & $-23808(8)^{~}$        & $2848(28)^{*~}$             & $-20960(27)^{*~}$ \\
		$^{44}$Ti  & $~0$        & $-37548.6(7)^{~}$      & $9330(4)^{**}$              & $-28218.6(40)^{**}$ \\
		$^{44}$Sc  & $+1$        & $-37816.0(18)^{~}$     & $2778(3)^{~~}$              & $-35038.2(25)^{~~}$ \\
		$^{44}$Ca  & $+2$        & $-41468.7(3)^{~}$      & $0^{~~}$                    & $-41468.7(3)^{~~}$ \\
		\hline
		\multicolumn{5}{l}{$^{*}$ from this work and $^{**}$ unperturbed level from Ref.~\cite{Dixon78}}\\
	\end{tabular*}
	\label{IMME table}
\end{table}

\begin{table*}[htbp]
	\caption{
		The coefficients obtained from the fitting by using quadratic, cubic, and quartic forms of the IMME.
		The corresponding mass data are given in Table~\ref{IMME table}.
	}
	\begin{ruledtabular}
		\begin{tabular}{cccccc}
			a              &      b            &       c        &      d    &     e        &  $\chi_{n}$ \\
			\hline
			$-28216.8\pm3.5 $ &  $-7018.1\pm 6.1$ & $ 196.1\pm 2.5 $ &           &              &  1.12 \\
			$-28217.3 \pm 3.9 $ & $ -7018.2 \pm 6.1 $ &  197.5 $\pm$5.9  & $-0.7 \pm 2.4$ &     &  1.49 \\
			$-28218.6 \pm 4.0 $ & $-7024.5 \pm 9.3 $ & 207.0 $\pm$12.2 &           & $-1.8 \pm 2.0 $  &  1.20 \\
			$-28218.6 \pm 4.0 $ & $-7048.2 \pm 18.6 $ & 228.2$\pm$18.9 & 9.1$\pm$6.2& $-8.7 \pm 5.1 $ &       \\
			
		\end{tabular}
	\end{ruledtabular}
	\label{IMME_fit}
\end{table*}

The $T=2$, $J^{\pi}=0^{+}$ IAS in $^{44}$V~\cite{Dossat} was originally proposed on the basis of $\beta$-delayed protons ($\beta$-p) from $^{44}$Cr decay.
There, the strongest proton branch with the relevant centre-of-mass energy, $E_p=910(11)$ keV, was assigned as decaying from the expected $T=2$, $J^{\pi}=0^{+}$ IAS in $^{44}$V to the ground state of $^{43}$Ti.
However, the branching ratio of this transition is only 1.7(3)\%, which is much smaller than the theoretical estimation of 28\% for the super-allowed $\beta$ decay of $^{44}$Cr to the $T=2$ IAS in $^{44}$V~\cite{Dossat}.
As no $\gamma$ transitions de-exciting the proposed IAS were observed to balance the $\beta$-feeding branching ratio, this assignment shall be carefully checked.

Recently, highly-sensitive measurements of $\beta$-delayed protons from $^{44}$Cr were conducted by employing an optical time projection chamber (OTPC), leading to the observation of a low-energy proton peak with a mean centre-of-mass energy of 759(26) keV~\cite{TPC}.
The ground-state mass of $^{44}$V has been measured at the CSRe~\cite{Regular_article_Zhang} and by Penning-trap mass spectrometry~\cite{LEBIT_44V}.
Meanwhile, detailed level structure of $^{44}$Sc including the Gamow-Teller transition strengths, B(GT), is available~\cite{Fujita13}.
These data enabled us to revisit the decay scheme of $^{44}$Cr and address the assignment of the $T=2$ IAS in $^{44}$V.

Table~\ref{summary of proton and gamma} summarises the available experimental information on $\beta$-delayed protons and $\gamma$-transitions from $^{44}$Cr decay.
All values are weighted-averages of the data from two measurements~\cite{Dossat,TPC}.
Given on the right side of Fig.~\ref{decay scheme} are the energy levels of $^{44}$Sc with B(GT)-values
larger than 0.1 as obtained in the $^{44}$Ca($^3$He,t)$^{44}$Sc reaction~\cite{Fujita13}.
Except for the IAS at $E_x=2779$ keV, other levels with $\Delta L=0$~\cite{Fujita13} are supposed to be $1^+$ states.
The weighted-average mass of $^{44}$V from Refs.~\cite{Regular_article_Zhang,LEBIT_44V} is used, giving the proton separation energy of $S_p(^{44}$V$)=1776(10)$ keV.

An interesting observation in the decay of $pf$-shell nuclei was made that the proton decay branches from the IAS in $^{53}$Co~\cite{2016Su10} and $^{52}$Co~\cite{2016XU10} are very weak or even non-observable.
Therefore, the conventional way to assign the strongest proton peak at a relevant centre-of-mass energy as
being from the IAS may cause misidentification.
The low or non-observable proton-decay branch has been attributed to
the very small isospin mixing of the related IAS~\cite{2016XU10}.
In the case of the proton decay of the IAS in $^{44}$V, the decay energy is expected to be less that 1 MeV, which is even smaller than in the cases of $^{53}$Co~\cite{2016Su10} and $^{52}$Co~\cite{2016XU10}.
As a consequence, the barrier penetration could be more difficult, providing higher hindrance for the proton decay of the IAS.
Furthermore, the assignment of the strongest 910-keV protons as being from the decay of the IAS in $^{44}$V to the ground state of $^{43}$Ti
is not supported by dedicated theoretical calculations~\cite{Smirnova}.
Two scenarios are proposed to understand the $\beta$-delayed proton emissions in $^{44}$Cr decay, which refer to the symmetry of the level structure in the mirror nuclei $^{44}$V and $^{44}$Sc.

First, it is assumed that no protons from the decay of the $T=2$ IAS in $^{44}$V were observed in the two experiments~\cite{Dossat,LEBIT_44V}.
For this case, a partial decay scheme of $^{44}$Cr is proposed and shown in Fig.~\ref{decay scheme}.
The strongest 910-keV proton decay branch is assigned to be the $^{44}{\rm V}(1^+,E_x=3161)\rightarrow $$^{43}{\rm Ti}(3/2^-,E_x=475)$ transition
rather than the $^{44}{\rm V}({\rm IAS},E_x=2686)\rightarrow$$^{43}{\rm Ti}(7/2^-,{\rm g.s.})$ transition suggested in Ref.~\cite{Dossat}.
We have estimated the log $ft$ values~\cite{ft} for each individual $\beta$ transition to the levels in $^{44}$V by using the experimental $\beta$ feedings, i.e. $I_{\beta}=I_p$.
The B(GT) strengths were determined from the $ft$ values via the equation
\begin{equation}\label{B(GT)}
B_{j}(GT)=\frac{K}{\lambda^{2}}\frac{1}{f_{j}t},
\end{equation}
where $K=6143.6(17)$~\cite{PPNP_Fujita}, $\lambda=-1.2701(25)$~\cite{Particle_Data},
the index $j$ represents the daughter state at the excitation energy $E_{j}$.
One sees that the level structure of $^{44}$V, including the level spacings and the deduced B(GT) strengths, is very similar to the analog states in $^{44}$Sc. By adopting the assignment of the 910-keV protons discussed above, the experimental $\beta$-delayed proton spectrum in Ref.~\cite{Dossat} agrees well with the $\beta$-decay spectrum deduced from the $^{44}$Ca($^3$He,t)$^{44}$Sc reaction~\cite{Fujita13}.

Second possibility is to assign the 759-keV protons as being from the $^{44}{\rm V}({\rm IAS},E_x=2848(28))\rightarrow$$^{43}{\rm Ti}(3/2^+,E_x=313)$ transition.
If referring to the mirror symmetry, this assignment gives more relevant excitation energies of the two IAS in $^{44}$V and $^{44}$Sc (see Fig.~\ref{decay scheme}).
This assignment can be checked with the IMME.
Table~\ref{IMME table} presents the $ME$-values of the $T=2$, $A=44$ quintet.
The mass data are fitted as in Refs.~\cite{2014MA56,2013Lam} using quadratic, cubic, and quartic forms of the IMME, and the obtained coefficients are listed in Table~\ref{IMME_fit}.
The $d$- and $e$-coefficients are compatible with zero within $2\sigma$.
This result indicates that the quadratic form of IMME is still valid if the location of the IAS in $^{44}$V as proposed here is adopted.
The large $d$- and $e$-coefficients demonstrated in Fig.~\ref{d coefficients}, the so-called breakdown of the IMME, are caused most probably by the misidentification of the IAS in $^{44}$V~\cite{Dossat}.
To confirm this conclusion, precision determination of the IAS through measurements of $\beta$-delayed $\gamma$ emissions in $^{44}$Cr decay is highly desired.

\section{Summary}\label{summary}
Mass measurements of neutron-deficient $fp$-shell nuclei produced in the projectile fragmentation of 468~MeV/u $^{58}$Ni beam were performed by using the isochronous mass spectrometry at the cooler storage ring CSRe in Lanzhou.
The masses of $^{44}$Cr, $^{46}$Mn, $^{48}$Fe, $^{50}$Co, and $^{52}$Ni were measured for the first time with relative precisions of $(1-2)\times$10$^{-6}$, and the mass precision for $^{40}$Ti was improved by a factor of 2.
The new mass values were compared with predictions of global mass models as well as local mass relations.
It is found that the experimental masses can be well described by the WS4 model with the radial basis function correction~\cite{Wang14}.
A systematic deviation seems to exist if comparing to the predictions of the Garvey-Kelson mass relation,
while good agreement was achieved with the IMME and the recent mirror-nuclei approach~\cite{Zong20}.

By using the new mass values, experimental data for five $T=2$ isospin quintets were completed including the heaviest one in the $pf$-shell,
namely for $A=44,46,50,{\rm and}~52$ quintets.
The extracted $d$- and $e$-coefficients of the quartic form of the IMME are compatible with zero within $2\sigma$ except for $A=44$.
The unexpectedly large $d$- and $e$-values were addressed by revisiting the experimental data on $\beta$-delayed protons from $^{44}$Cr decay.
It is suggested that the strongest 910-keV proton branch is not from the de-excitation of the IAS in $^{44}$V.
It is concluded that the observed breakdown of the quadratic form of the IMME as well as the large $d$- and $e$-coefficients exhibited in Fig.~\ref{d coefficients} most probably originate from the misidentification of the IAS in $^{44}$V~\cite{Dossat}.
To confirm this conclusion, precision determination of the IAS through measurements of $\beta$-delayed $\gamma$ emissions in $^{44}$Cr decay is highly desired.

\acknowledgments
We thank the staff of the accelerator division of the IMP for providing the stable beam.
This work was supported in part by the National Key R\&D Program of China (Grant No. 2018YFA0404400, No. 2016YFA0400504 and No. 2016YFA0400501), the Strategic Priority Research Program of Chinese Academy of Sciences (Grant No. XDB34000000),
the Key Research Program of Frontier Sciences of CAS (Grant No. QYZDJ-SSW-S), the NSFC grants 11905259, 11905261, 11975280, U1932206, 11805032, 11775277, 11961141004, and the Helmholtz-CAS Joint Research Group HCJRG-108.
Y.A.L. acknowledges support from European Research Council (ERC) under the EU Horizon 2020 Research and Innovation Programme (Grant Agreement No. 682841 "ASTRUm"). T.U., T.Y., and A.O. are supported in part by JSPS and NSFC under the "Japan-China Scientific Cooperation Program". C.Y.F. and Y.M.X. are thankful for the support from CAS "Light of West China" Program.

\end{document}